\documentclass[sigconf]{acmart}

\AtBeginDocument{%
  }

\usepackage{subcaption}
\usepackage{listings}
\definecolor{codegreen}{rgb}{0,0.6,0}
\definecolor{codegray}{rgb}{0.5,0.5,0.5}
\definecolor{codepurple}{rgb}{0.58,0,0.82}
\definecolor{backcolour}{rgb}{0.95,0.95,0.92}

\lstdefinestyle{mystyle}{
    commentstyle=\color{codegreen},
    keywordstyle=\color{magenta},
    numberstyle=\tiny\color{codegray},
    stringstyle=\color{codepurple},
    basicstyle=\ttfamily\footnotesize,
    breakatwhitespace=false,         
    breaklines=true,                 
    captionpos=b,                    
    keepspaces=true,                 
    numbers=left,                    
    numbersep=5pt,                  
    showspaces=false,                
    showstringspaces=false,
    showtabs=false,                  
    tabsize=2
}

\copyrightyear{2026}
\acmYear{2026}
\setcopyright{cc}
\setcctype{by}
\acmConference[MSR '26]{23rd International Conference on Mining Software Repositories}{April 13--14, 2026}{Rio de Janeiro, Brazil}
\acmBooktitle{23rd International Conference on Mining Software Repositories (MSR '26), April 13--14, 2026, Rio de Janeiro, Brazil}
\acmPrice{}
\acmDOI{10.1145/3793302.3793327}
\acmISBN{979-8-4007-2474-9/2026/04}

\begin{document}

\title{GitEvo: Code Evolution Analysis for Git Repositories}

\author{Andre Hora}
\orcid{0000-0003-4900-1330}
\affiliation{%
  \institution{Department of Computer Science, UFMG}
  \city{Belo Horizonte}
  \country{Brazil}
}
\email{andrehora@dcc.ufmg.br}

\begin{abstract}
Analyzing the code evolution of software systems is relevant for practitioners, researchers, and educators.
It can help practitioners identify design trends and maintenance challenges, provide researchers with empirical data to study changes over time, and give educators real-world examples that enhance the teaching of software evolution concepts.
Unfortunately, we lack tools specifically designed to support code evolution analysis.
In this paper, we propose GitEvo, a multi-language and extensible tool for analyzing code evolution in Git repositories.
GitEvo leverages Git frameworks and code parsing tools to integrate both Git-level and code-level analysis.
We conclude by describing how GitEvo can support the development of novel empirical studies on code evolution and act as a learning tool for educators and students.
GitEvo is available at: \url{https://github.com/andrehora/gitevo}.
\end{abstract}

%
%




\maketitle

\section{Introduction}

Software evolution assessment provides multiple benefits for practitioners, researchers, and educators~\cite{lehman1996laws, lehman2006software, mens2005challenges, brito2020refactoring, brito2022understanding, hora2018assessing}.
Practitioners can identify design trends and potential maintenance challenges.
Researchers can assess empirical data to investigate how software evolves over time.
Educators can draw on real-world examples to enrich the teaching of software evolution concepts.

Nowadays, most real-world software systems are managed using the version control system Git~\cite{git, so_survey}.
Thus, a common approach to exploring software evolution is to rely on tools that analyze Git repositories, such as GitPython~\cite{gitpython} and PyDriller~\cite{pydriller, spadini2018pydriller} for Python, JGit~\cite{jgit} for Java, and isomorphic-git~\cite{isomorphic-git} for JavaScript, to name a few.
These Git-level tools allow practitioners to assess repository changes over time, such as commits, authors, and file modifications.
However, these tools do not support code-level analysis; they are limited to extracting Git-level data.
As a result, they cannot provide insights into how the \emph{source code} itself evolves.

\begin{figure}[t]
     \centering
     \begin{subfigure}[b]{0.23\textwidth}
         \centering
         \includegraphics[width=\textwidth]{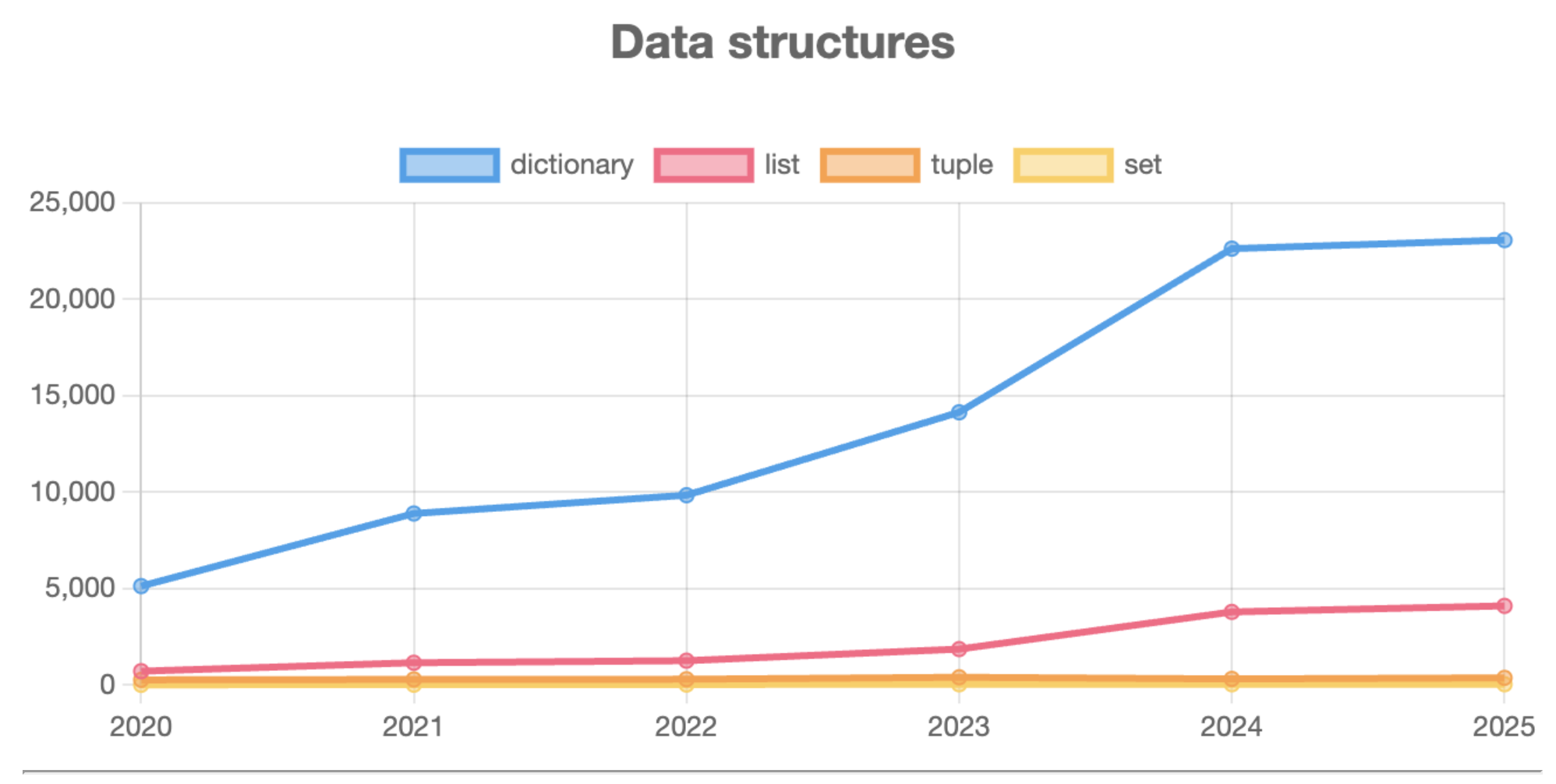}
         \caption{Python - FastAPI}
         \label{fig:1a}
     \end{subfigure}
     \begin{subfigure}[b]{0.23\textwidth}
         \centering
         \includegraphics[width=\textwidth]{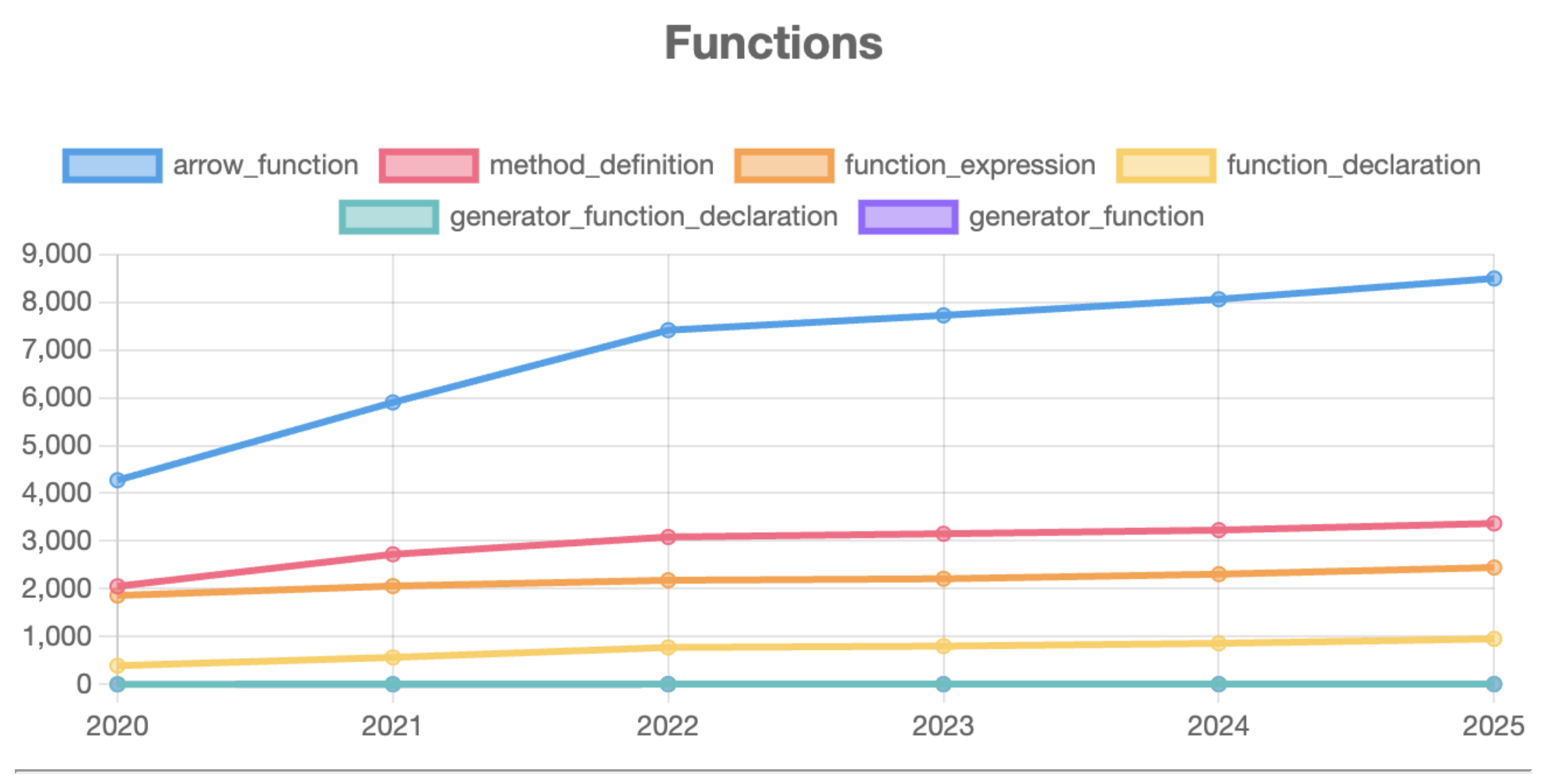}
         \caption{JavaScript - Webpack}
         \label{fig:1b}
     \end{subfigure}
     \begin{subfigure}[b]{0.23\textwidth}
         \centering
         \includegraphics[width=\textwidth]{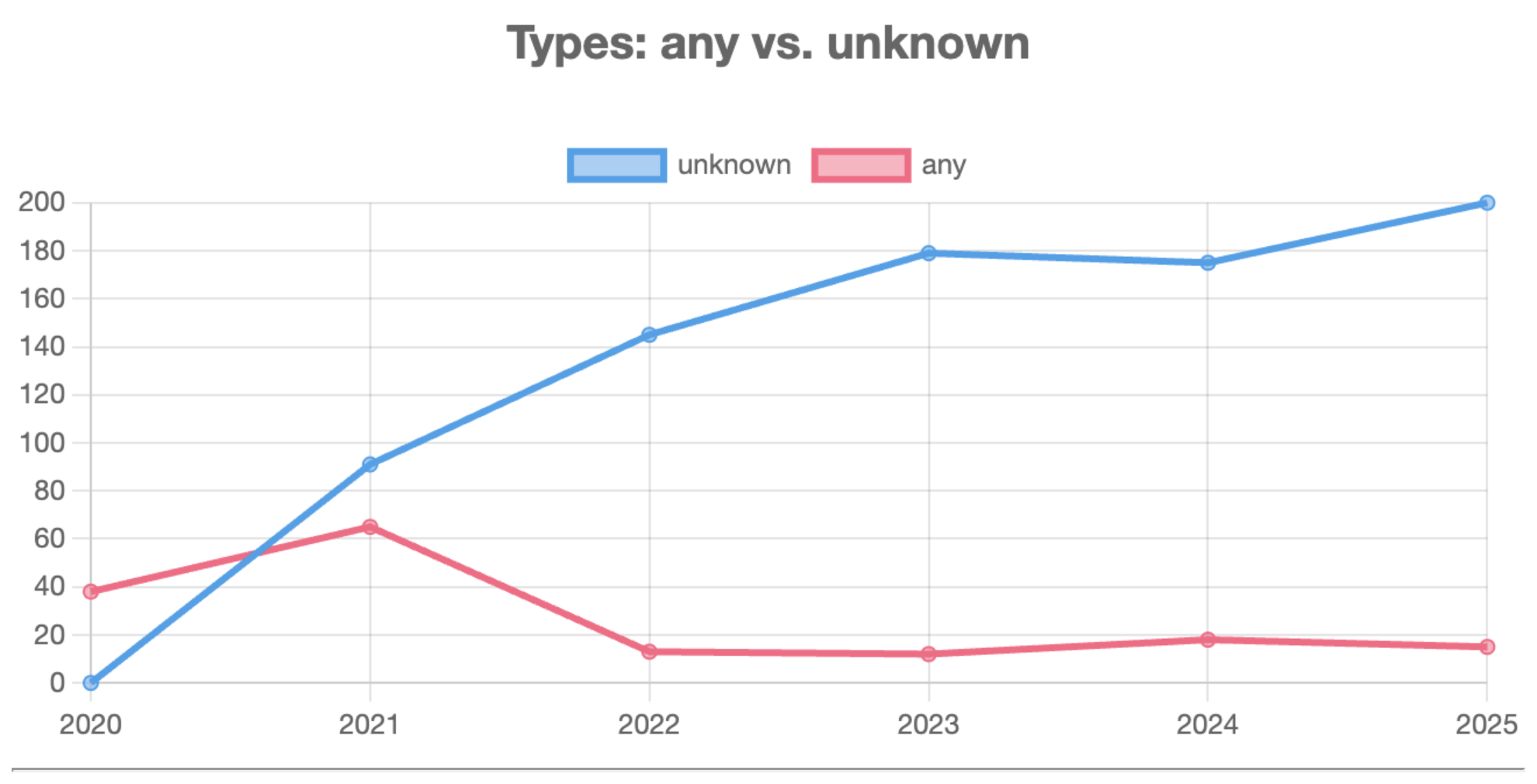}
         \caption{TypeScript - Docusaurus}
         \label{fig:1c}
     \end{subfigure}
     \begin{subfigure}[b]{0.23\textwidth}
         \centering
         \includegraphics[width=\textwidth]{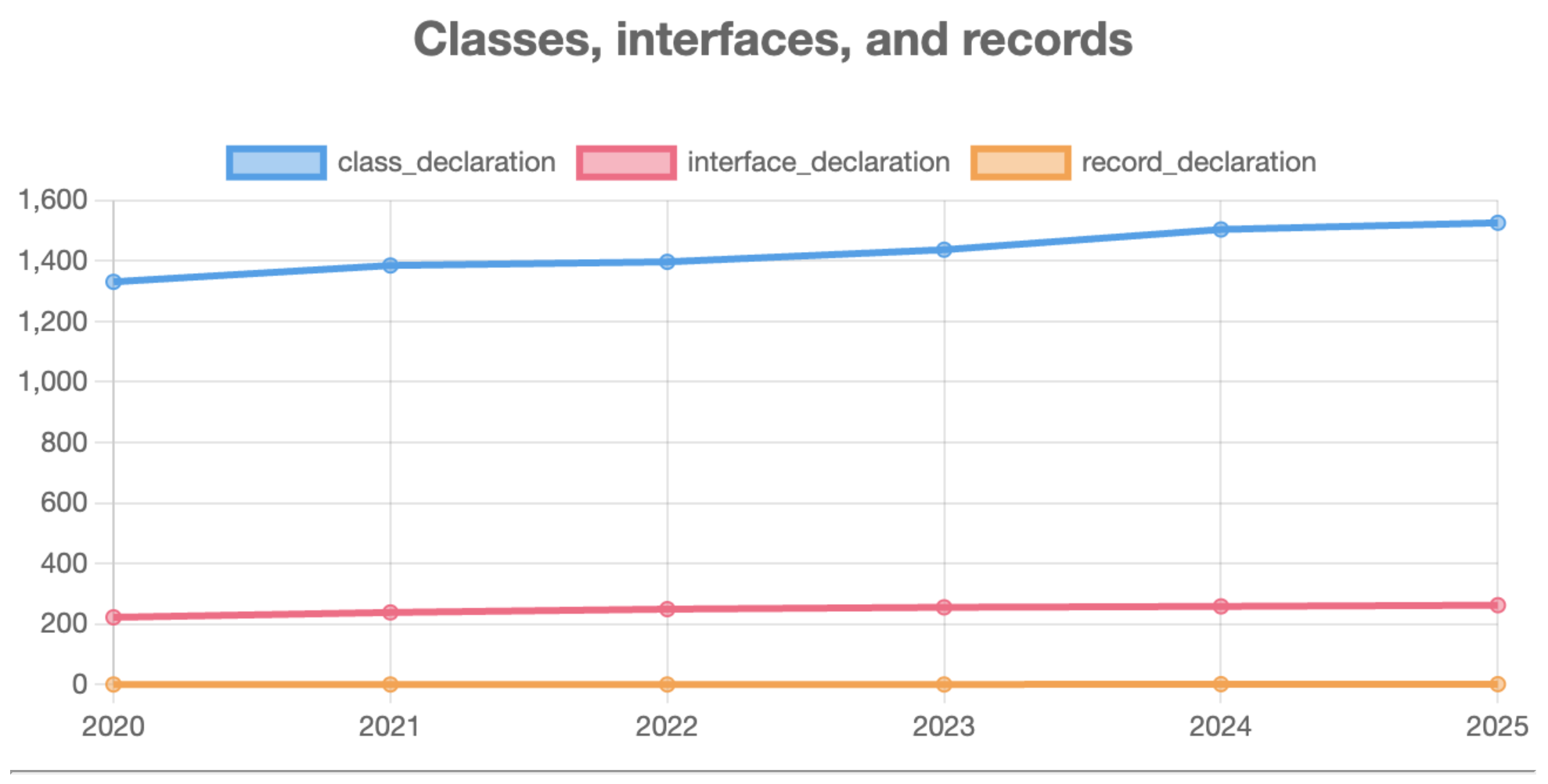}
         \caption{Java - Mockito}
         \label{fig:1d}
     \end{subfigure}
    \caption{Examples of code evolution reports by GitEvo.}
    \Description{Examples of code evolution reports by GitEvo.}
    \label{fig:examples}
\end{figure}

On the other hand, code-level analysis can be performed using code parsing tools.
There is a wide variety of such tools available.
Some tools are more generic and focus on processing syntax trees, such as the AST module~\cite{python-ast} for Python, JavaParser~\cite{javaparser} and Spoon~\cite{java-spoon, spoon-paper} for Java, Babel parser~\cite{js-babel-parser} for JavaScript, and Tree-sitter~\cite{tree-sitter} for multiple programming languages.
Many other tools serve more specific purposes, for example, computing code complexity~\cite{lizard}, LOC~\cite{cloc}, metrics~\cite{ck-aniche}, and code changes~\cite{german2019cregit, narkebski2025patchscope}.
However, these code-level tools operate independently of the Git version control system and are therefore limited to analyzing isolated code snapshots or individual files, rather than capturing the evolution of code over time.

This scenario brings to light two main issues.
\emph{First}, there is a lack of tools specifically designed to support code evolution analysis, that is, tools that combine both Git-level \emph{and} code-level assessment.
\emph{Second}, existing solutions are often language-dependent.
For example, to analyze the evolution of Python code, one might use GitPython together with the AST module.
However, performing a similar analysis for another language, such as Java, would require re-implementing the solution using JGit and Spoon.

\begin{figure*}[t]
     \centering
         \includegraphics[width=0.8\textwidth]{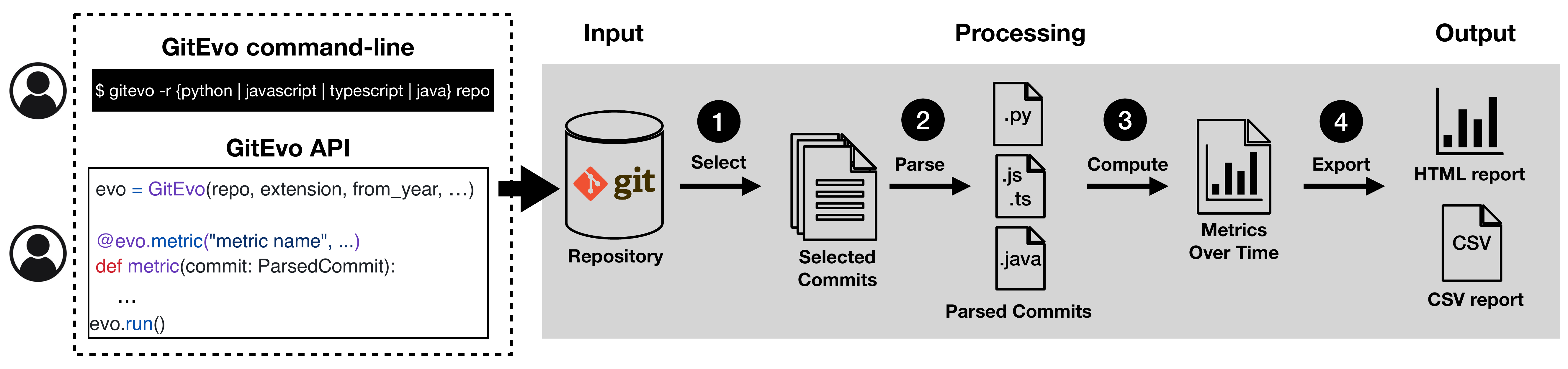}
         \caption{Overview of the GitEvo pipeline.}
         \Description{Overview of the GitEvo pipeline.}
        \label{fig:gitevo}
\end{figure*}


We propose GitEvo, a multi-language and extensible Python-based tool for analyzing code evolution in Git repositories.\footnote{\url{https://github.com/andrehora/gitevo}}
GitEvo currently supports the analysis of Python, JavaScript, TypeScript, and Java.
It leverages Git and code parsing tools to integrate Git-level and code-level analysis.
Figure~\ref{fig:examples} presents reports generated by GitEvo to analyze code evolution in multiple languages.

GitEvo relies on GitPython~\cite{gitpython} and PyDriller~\cite{spadini2018pydriller} for Git-level analysis, and on Tree-sitter~\cite{tree-sitter} for multi-language, code-level assessment (Section~\ref{fig:gitevo}).
It enables users to write \emph{scripts} entirely in Python to compute code evolution metrics at the level of the concrete syntax tree (CST), leveraging Tree-sitter for parsing.
For example, users can extract evolution metrics about various program constructs, such as classes, methods, functions, decorators, variables, parameters, types, and data structures, to name a few.

Finally, we showcase practical applications of GitEvo (Section~\ref{sec:applications}).
For instance, we describe how we employed GitEvo in two large-scale empirical studies and in an undergraduate Software Engineering course to support over 90 students exploring the evolution of real-world software systems.

\textbf{Novelty.}
GitEvo’s novelty lies in its ability to combine both Git-level and code-level analyses within a single tool, reducing the need for multiple tools or programming languages.
In our experience, this approach reduces effort, allowing users to focus on what information needs to be analyzed over time rather than on which technologies should be adopted.

\section{GitEvo}
\label{sec:gitevo}

\subsection{Overview}

Figure~\ref{fig:gitevo} provides an overview of the GitEvo pipeline.
GitEvo can be used from the command-line (Section~\ref{sec:cli}) or programmatically via API (Section~\ref{sec:api}).
When using via API, GitEvo is extensible, and users can define custom metrics.
GitEvo receives as input one or more Git repositories and provides as output HTML and CSV reports of the code-level metrics over time.
As an example, we make dozens of HTML reports available in \texttt{gitevo-examples}.\footnote{\url{https://github.com/andrehora/gitevo-examples}}

\subsection{Usage via Command-Line}
\label{sec:cli}

The simplest way to use GitEvo is via the command-line.
After installing GitEvo, we can analyze the evolution of a Git repository:

\begin{lstlisting}[language=Bash, backgroundcolor=\color{backcolour}, numbers=none]
# Installing GitEvo
$ pip install gitevo
# Basic command-line usage
$ gitevo -r {python|javascript|typescript|java} repo
\end{lstlisting}

\texttt{repo} accepts (1) a Git URL, (2) a local repository, or (3) a directory containing multiple Git repositories.
GitEvo provides as output HTML and CSV reports.

For example, the following commands analyze the source code evolution of 
Flask (Python),\footnote{\url{https://andrehora.github.io/gitevo-examples/python/flask.html}}
Axios (JavaScript),\footnote{\url{https://andrehora.github.io/gitevo-examples/javascript/axios.html}}
VueJS-core (TypeScript),\footnote{\url{https://andrehora.github.io/gitevo-examples/typescript/vuejs-core.html}} and
Mockito (Java).\footnote{\url{https://andrehora.github.io/gitevo-examples/java/mockito.html}}
The resulting HTML reports are all available in our \texttt{gitevo-examples} repository.

\begin{lstlisting}[language=Bash, backgroundcolor=\color{backcolour}, numbers=none]
# Using GitEvo to analyze Python, JS, TS, and Java code
$ gitevo -r python https://github.com/pallets/flask
$ gitevo -r javascript https://github.com/axios/axios
$ gitevo -r typescript https://github.com/vuejs/core
$ gitevo -r java https://github.com/mockito/mockito
\end{lstlisting}

When using GitEvo via the command-line, the computed code metrics are predefined within the tool.\footnote{\url{https://github.com/andrehora/gitevo/tree/main/gitevo/reports}}
For example, for Python, GitEvo computes lines of code, data structures, function parameters, and many more.
To define custom metrics, users can use GitEvo through its API, as described next.

\subsection{Usage via API: Defining Custom Metrics}
\label{sec:api}

\noindent\textbf{Overview.}
The most powerful way to use GitEvo is programmatically through its API.
In this mode, users can implement Python scripts to define custom code evolution metrics at the level of the concrete syntax tree (CST).
These scripts are executed by GitEvo, which automatically generates HTML and CSV reports.
Behind the scenes, the script performs four major steps:
(1) selects representative commits,
(2) parses source files,
(3) computes metrics, and
(4) export reports, as illustrated in Figure~\ref{fig:gitevo}.

\begin{figure}[h]
     \centering
         \fbox{\includegraphics[width=0.42\textwidth]{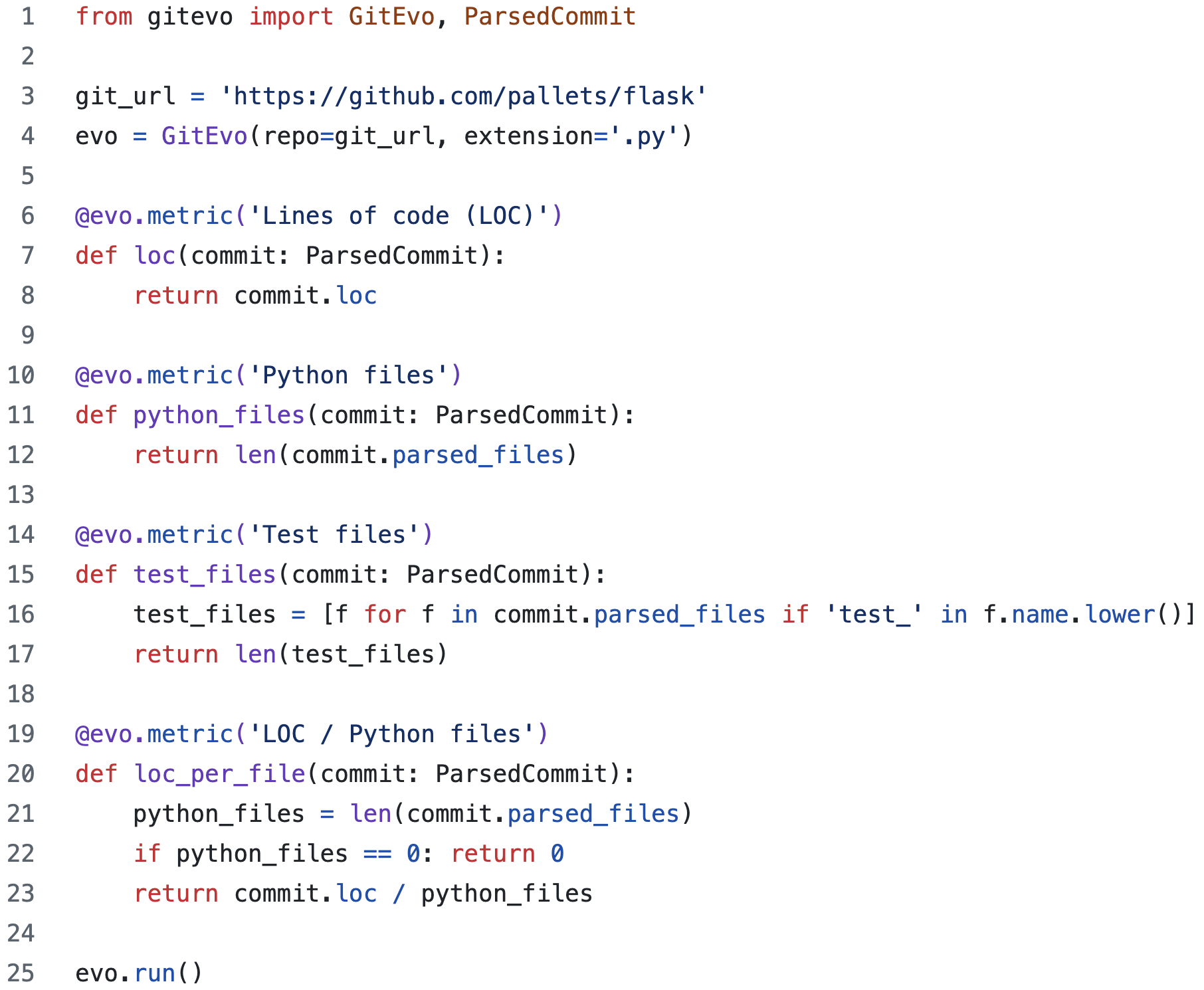}}
         \caption{Example of a basic GitEvo script.}
         \Description{Example of a basic GitEvo script.}
        \label{fig:gitevo-ex1}
\end{figure}

\begin{figure}[h]
     \centering
         \includegraphics[width=0.45\textwidth]{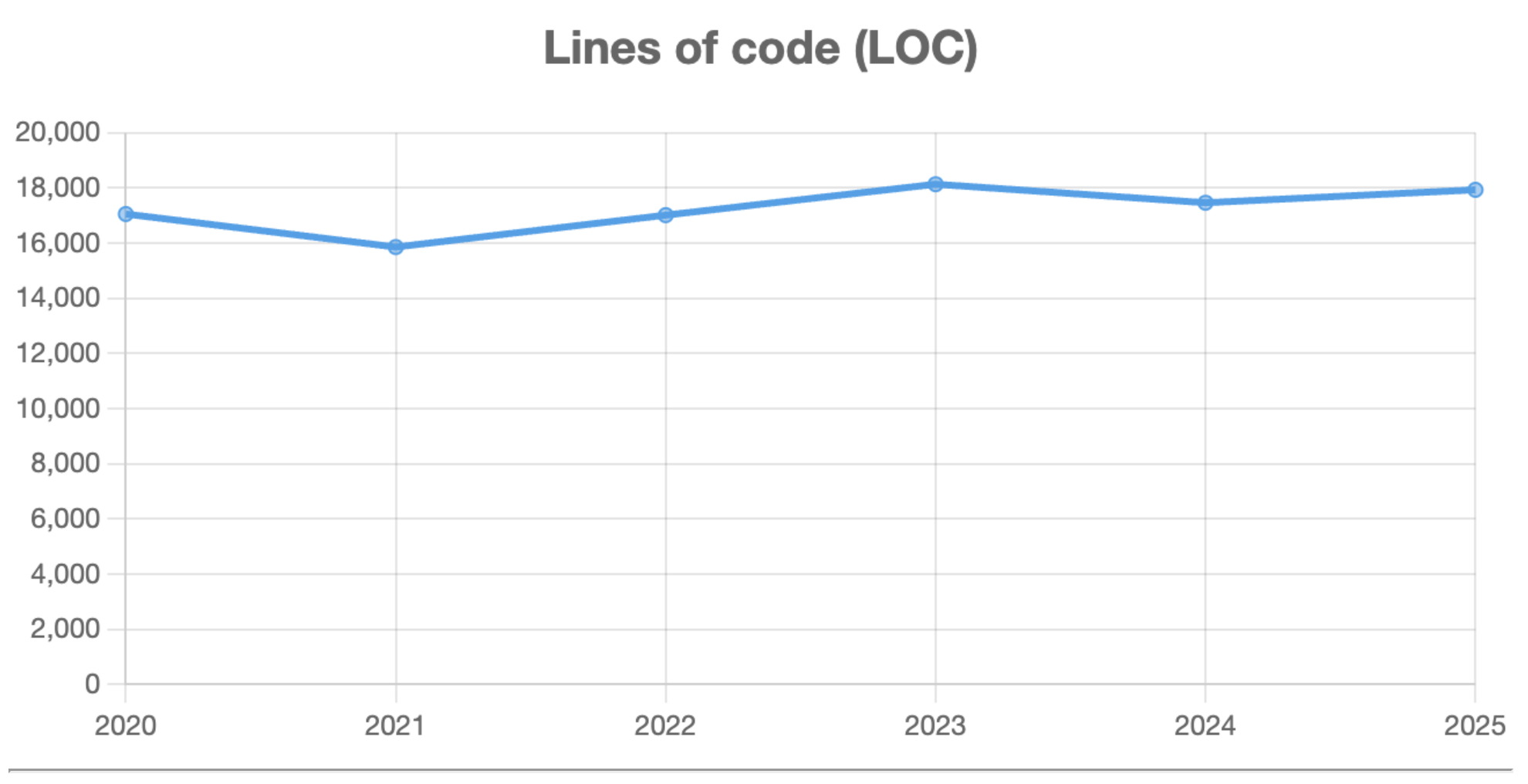}
         \caption{Lines of code over time (Flask).}
         \Description{Lines of code over time (Flask).}
        \label{fig:flask-loc}
\end{figure}

\noindent\textbf{Example 1 (basic evolution metrics).}
Figure~\ref{fig:gitevo-ex1} presents an example of a GitEvo script that computes four metrics: Lines of Code, Python files, Test files, and LOC per Python file.\footnote{\url{https://github.com/andrehora/gitevo/blob/main/examples/ex_basic.py}}
The Lines of Code metric (lines 6–8) produces the chart shown in Figure~\ref{fig:flask-loc}.

\noindent\textbf{API.}
GitEvo provides three key classes that can be used in the scripts: \texttt{GitEvo}, \texttt{ParsedCommit}, and \texttt{ParsedFile}.
\texttt{GitEvo} is the main class, the entry point to use the tool.
It receives as input the repository, file extension, date unit for analysis, and start/end year for analysis.
GitEvo can analyze code evolution yearly (first day of each year) and monthly (first day of each month).
By default, GitEvo analyzes the last five years.

Metrics are defined in functions decorated with \texttt{@evo.metric()}.
The metric decorator can include the metric name and other arguments to customize the metric output.
Notice that the metric function receives as an argument a \texttt{ParsedCommit}.

The \texttt{ParsedCommit} class represents a parsed commit and contains (1) a list of \texttt{ParsedFile} and (2) a list of \texttt{tree\_sitter.Node}.\footnote{\url{https://tree-sitter.github.io/py-tree-sitter/classes/tree_sitter.Node.html}}
\texttt{ParsedCommit} also contains multiple methods (e.g.,~\texttt{find\_\-node\_\-types} and \texttt{loc\_by\_type}) and properties (e.g.,~commit hash and loc).
A \texttt{ParsedFile} represents a parsed file in a commit, including properties as name, path, and tree-sitter nodes.
A \texttt{tree\_sitter.Node} is a single node within the CST, including type, parent, and children.

\begin{figure}[h]
     \centering
         \fbox{\includegraphics[width=0.46\textwidth]{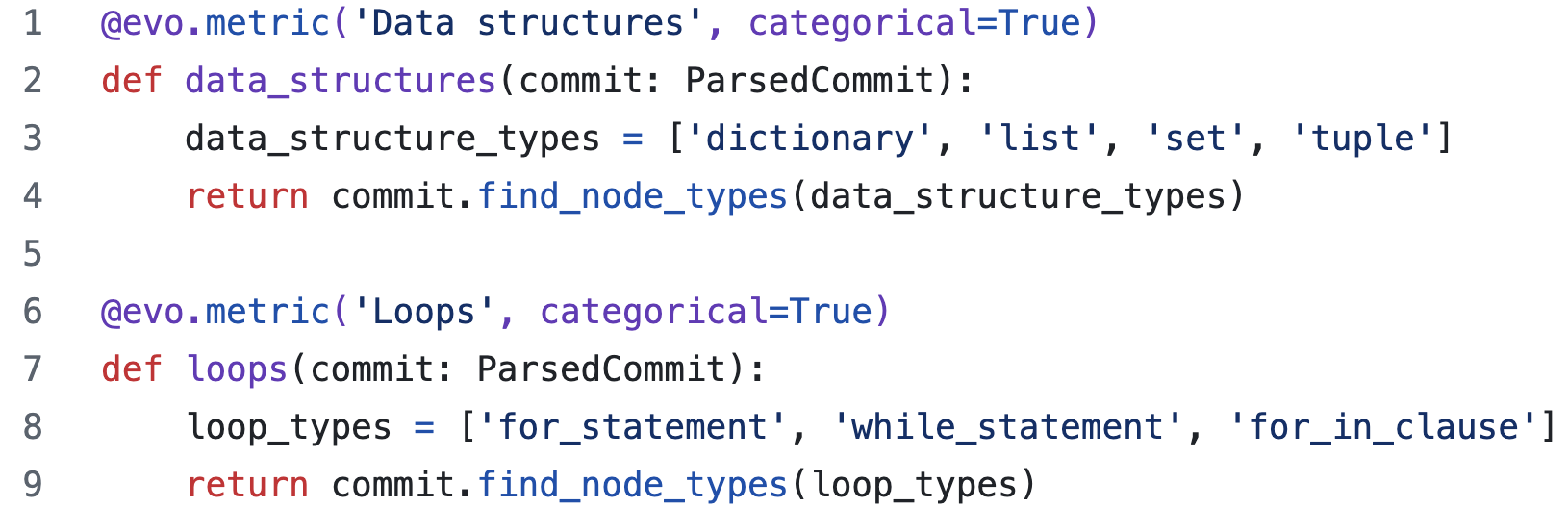}}
         \caption{Metrics based on node types.}
         \Description{Metrics based on node types.}
        \label{fig:gitevo-ex2}
\end{figure}

\begin{figure}[h]
     \centering
         \includegraphics[width=0.45\textwidth]{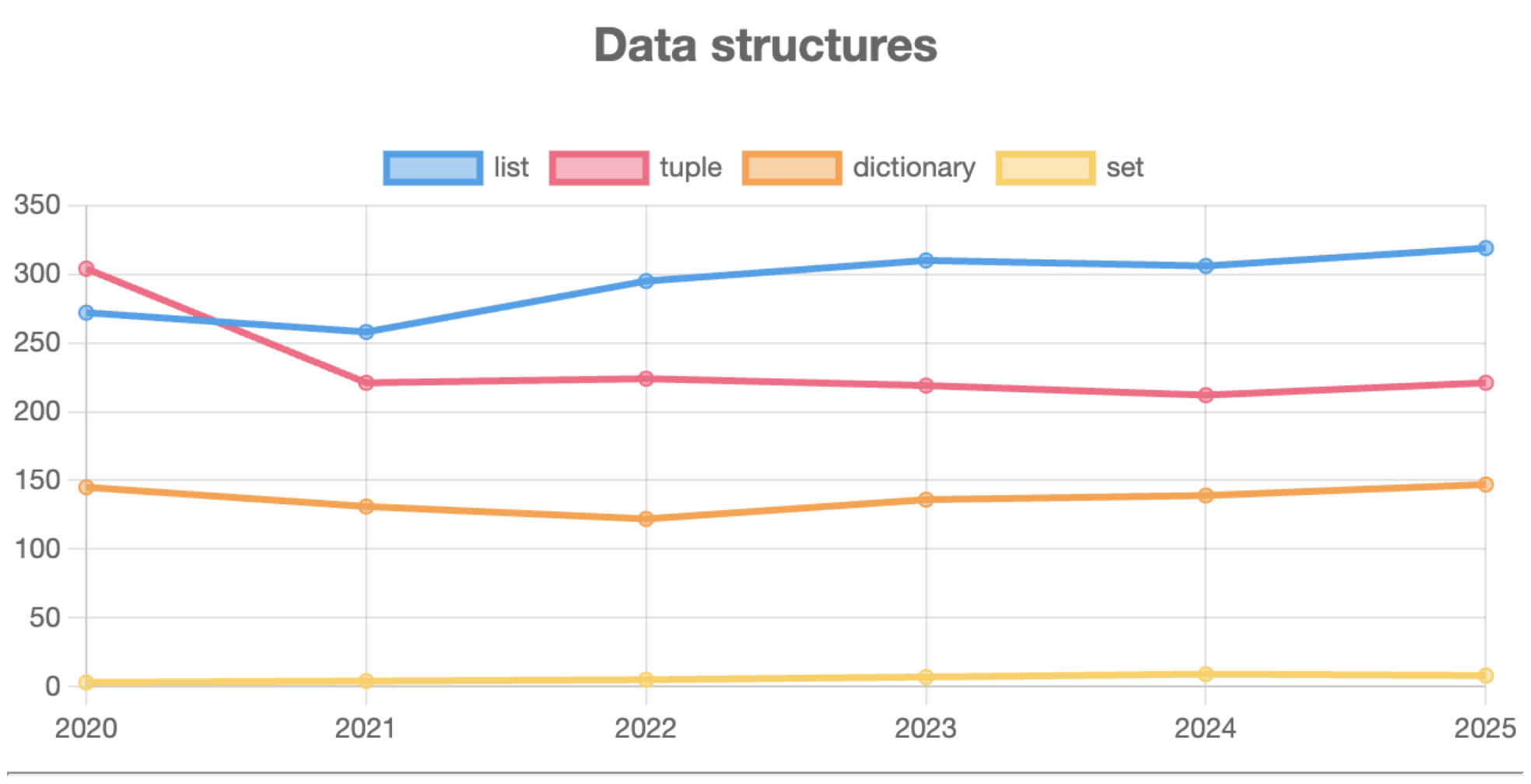}
         \caption{Data structures over time (Flask).}
         \Description{Data structures over time (Flask).}
        \label{fig:flask-data-struc}
\end{figure}

\noindent\textbf{Example 2 (metrics based on node types).}
Figure~\ref{fig:gitevo-ex2} presents two metrics (Data structures and Loop) that use \texttt{find\_\-node\_\-types} to count nodes by type.\footnote{\url{https://github.com/tree-sitter/tree-sitter-python/blob/master/src/node-types.json}}
The Data structures metric (lines 1–4) produces the chart shown in Figure~\ref{fig:flask-data-struc}.


\begin{figure}[h]
     \centering
         \includegraphics[width=0.45\textwidth]{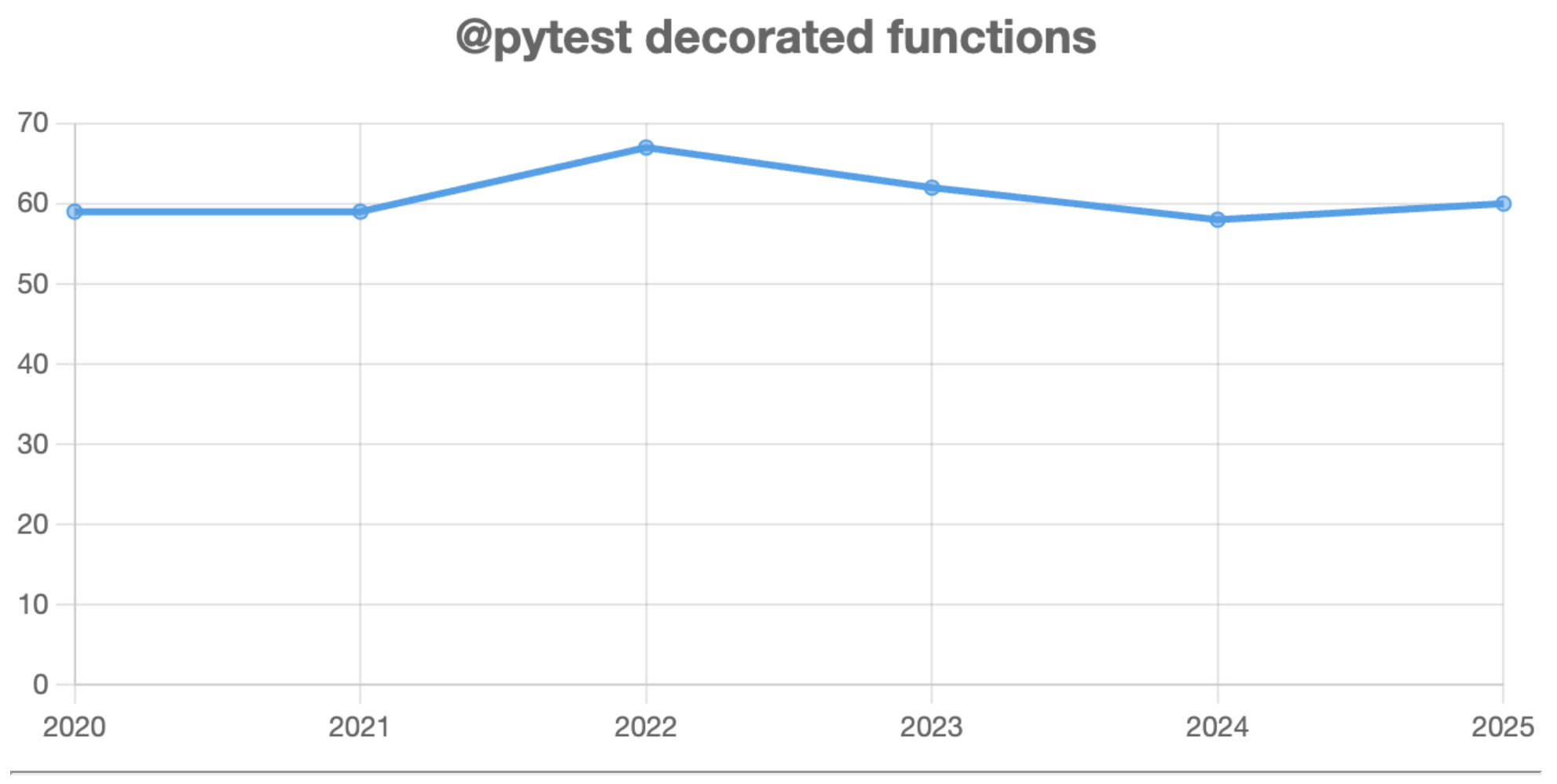}
         \caption{@pytest decorated functions over time (Flask).}
         \Description{@pytest decorated functions over time (Flask).}
        \label{fig:flask-pytest}
\end{figure}

\noindent\textbf{Example 3 (metrics based on node content).}
Figure~\ref{fig:gitevo-ex2} presents two metrics (Async functions and @pytest decorated functions) that are computed from the node content, leveraging the \texttt{tree\_sitter.\-Node} properties.
The ``@pytest decorated functions'' metric (lines 7–12) produces the chart shown in Figure~\ref{fig:flask-pytest}.

\begin{figure}[h]
     \centering
         \fbox{\includegraphics[width=0.46\textwidth]{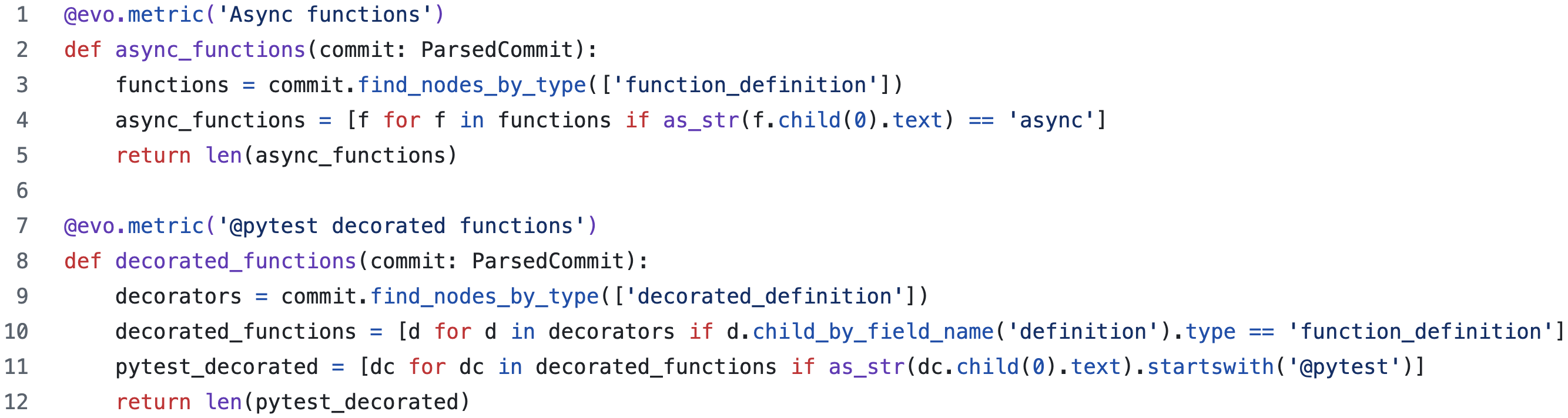}}
         \caption{Metrics based on node content.}
         \Description{Metrics based on node content.}
        \label{fig:gitevo-ex3}
\end{figure}

\subsection{Implementation Notes}

GitEvo leverages GitPython~\cite{gitpython} and PyDriller~\cite{spadini2018pydriller} for Git-level analysis, and Tree-sitter~\cite{tree-sitter} for multi-language, code-level assessment.
Behind the scenes, GitEvo orchestrates the analysis by creating a \texttt{ParsedCommit} object for each analyzed commit and injecting this object into the metric function.
This way, the user simply defines each metric once (in the metric function), and its value is automatically computed over time across all analyzed commits.

\noindent\textbf{GitPython and PyDriller.} GitEvo relies on GitPython and PyDriller to clone/open repositories, iterate over commits, and retrieve the source code in each commit. However, these tools do not perform code analysis. To analyze source code across multiple programming languages, a parser such as Tree-sitter is required.
    
\noindent\textbf{Tree-sitter.} GitEvo relies on Tree-sitter to parse the source code of each target commit. GitEvo identifies the target programming language and parses the source code using the corresponding Tree-sitter grammar.
The parsed code can be accessed by the user through the classes \texttt{ParsedCommit} and \texttt{ParsedFile}.
Currently, GitEvo supports four grammars: tree-sitter-python~\cite{tree-sitter-python}, tree-sitter-js~\cite{tree-sitter-javascript}, tree-sitter-ts~\cite{tree-sitter-typescript}, and tree-sitter-java~\cite{tree-sitter-java}.

\section{Practical Applications}
\label{sec:applications}

\subsection{Empirical Studies and Datasets}

GitEvo can support the development of novel studies on code evolution across multiple programming languages.
This allows researchers to analyze repositories written in different languages while writing analysis scripts in Python.
As an example, we detail two large-scale empirical studies in which we employed GitEvo.

\noindent\textbf{Mocking practices over time.}
Developers can use mocking in tests to isolate dependencies, making the test fast and repeatable~\cite{meszaros2007xunit, pereira2020assessing}.
We analyzed more than 1.2 million commits across 2,168 TypeScript, JavaScript, and Python repositories, identifying 44,900 commits that add mocks to tests (see Table~\ref{tab:summary-repos-commits}).
Although multiple programming languages were analyzed, all scripts were implemented in Python itself with the support of GitEvo.

\begin{table}[h]
    \centering
    \small
    \caption{Dataset of commits adding mocks to tests.}
    \begin{tabular}{lrrrr}
        \toprule
        \textbf{Commits} & \textbf{TS} & \textbf{JS} & \textbf{Python} & \textbf{Total} \\ 
        \midrule
        All commits & 835,781 & 98,389 & 320,708 & 1,254,878 \\
        Mock commits & 23,838 & 1,561 & 19,501 & 44,900 \\
        \bottomrule
    \end{tabular}
    \label{tab:summary-repos-commits}
\end{table}


\noindent\textbf{Functional features over time.}
We assessed 10 years of functional Python feature usage across three major open-source projects: CPython, Pandas, and Django.
We analyzed six functional features: lambdas, yield statements, generator expressions, list comprehensions, dictionary comprehensions, and set comprehensions~\cite{yang2022complex}.
We found that \emph{lambdas} are the most frequently used functional feature, with 9,988 cases, followed by \emph{yield} statements (5,552) and \emph{list comprehensions} (5,130).
In contrast, generator expressions (2,454), dictionary comprehensions (657), and set comprehensions (288) are less frequently used.
The metrics were computed entirely using GitEvo scripts.\footnote{\url{https://github.com/andrehora/gitevo/blob/main/examples/python_functional.py}}


\emph{Practical Application 1}: 
GitEvo can support the development of novel empirical studies on code evolution across multiple programming languages.

\subsection{Custom Reports for Practitioners}

GitEvo can also be used by practitioners to monitor the trends of specific metrics in their repositories.
As an example, we created a custom report for the FastAPI web framework~\cite{fastapi}, which includes several metrics related to web development.\footnote{\url{https://andrehora.github.io/gitevo-examples/fastapi/dispatch.html}}

Moreover, we presented GitEvo to practitioners via GitHub Discussions and received several positive feedback comments.
For instance, one developer stated: ``\emph{really great to take a step back and see some of the work we have done at this level!}''.\footnote{\url{https://github.com/pypa/pipenv/discussions/6376}}
Other developers suggested new features,\footnote{\url{https://github.com/andrehora/gitevo/issues/1}} improvements,\footnote{\url{https://github.com/sveltejs/svelte/discussions/15784}} and clarifications.\footnote{\url{https://github.com/mkdocs/mkdocs/discussions/3966}}
Although GitEvo is still a prototype, these interactions demonstrate its potential to attract practitioners’ interest.

\emph{Practical Application 2}: 
GitEvo can help practitioners visualize and understand how their software evolves.

\subsection{Education in Software Evolution}

We envision GitEvo as a learning tool to help students gain a deeper understanding of software evolution.
In this context, we employed GitEvo in an undergraduate Software Engineering course to support students exploring the evolution of real-world software systems.
The exercise was conducted with over 90 students and is publicly available~\cite{exploring-code-evolution} (each \texttt{fork} corresponds to an individual student’s answer).
The exercise consisted of four main steps: 
(1) select a repository to analyze,
(2) install and run the GitEvo tool,
(3) explore the code evolution report, and
(4) explain a code evolution chart.

The students identified many explanations for the observed code evolution patterns by exploring project documentation and inspecting the source code.
For example, they detected real cases of increasing code complexity, rising proportion of tests, and decreasing ratio of code comments.
They also identified several project-specific design decisions, such as the preference for \texttt{const} over \texttt{var} and \texttt{let} (in JS) and the adoption of \texttt{record} (in Java), to name a few.


\emph{Practical Application 3}: 
GitEvo can serve as a learning tool for educators and students to gain deeper insights into real software evolution and its associated challenges.



\section{Conclusion}

In this paper, we presented GitEvo, a Python-based tool that analyzes the evolution of code metrics across Git repository history.
We also showcased practical applications of GitEvo for researchers, practitioners, and educators.

As future work, we plan to extend support to any programming language available in Tree-sitter, only requiring the user to install the proper grammar.
We also plan to enhance the GitEvo API by adding more convenient methods for accessing common source code constructs, as classes, methods, functions, and decorators.


\begin{acks}
This research was supported by CNPq (process 403304/2025-3), CAPES, and FAPEMIG.
This work was partially supported by INES.IA (National Institute of Science and Technology for Software Engineering Based on and for Artificial Intelligence), www.ines.org.br, CNPq grant 408817/2024-0.
\end{acks}

\bibliographystyle{ACM-Reference-Format}
\bibliography{main}

\end{document}